# Superparaelectric phase in the ensemble of non-interacting ferroelectric nanoparticles

M.D. Glinchuk, E.A. Eliseev, A.N. Morozovska,[*]

Institute for Problems of Materials Science, NAS of Ukraine,

Krjijanovskogo 3, 03142 Kiev, Ukraine

**Abstract**

For the first time we predict the conditions of superparaelectric phase appearance in the ensemble of non-interacting spherical ferroelectric nanoparticles. The superparaelectricity in nanoparticle was defined by analogy with superparamagnetism, obtained earlier in small nanoparticles made of paramagnetic material.

Calculations of correlation radius, energetic barriers of polarization reorientation and polarization response to external electric field, performed within Landau-Ginzburg phenomenological approach for perovskites Pb(Zr,Ti)O$_3$, BiFeO$_3$ and uniaxial ferroelectrics rochelle salt and triglycine sulfate, proved that under the favorable conditions ensemble of their noninteracting nanoparticles possesses the following characteristic features. Namely:

(1) When the particle radius $R$ is less than the correlation radius $R_c$, but higher than the critical radius $R_{cr}$ of size-driven ferroelectric-paraelectric phase transition, all dipole moments inside the particle are aligned due to the correlation effects.

(2) Potential barrier $\Delta F(T,R)$ of polarization reorientation can be smaller than thermal activation energy at temperatures $T$ higher than the freezing temperature $T_f(R)$ depending the particle radius $R$. Freezing temperature $T_f(R)$ can be estimated from the condition $\Delta F(T,R) = \gamma k_B T$. Such determination is somehow voluntary, since the rigorous value of $T_f(R)$ depends on numerical factor $\gamma$ before $k_B T$, which depends on the system characteristics.

(3) Langevin-like law for polarization dependence on external field is obtained at temperatures higher than the freezing temperature $T_f(R)$, but lower than the temperature $T_{cr}(R)$ of size-driven ferroelectric-paraelectric phase transition.

(4) Ferroelectric hysteresis loop and remnant polarization appear at temperatures $T$ lower than the freezing temperature $T_f(R)$. This behaviour can be named as frozen superparaelectric phase.

The conditions (1)-(4) determine the superparaelectric phase appearance in the ensemble of ferroelectric nanoparticles of radius $R_{cr} < R < R_c$ at temperatures $T_f(R) < T < T_{cr}(R)$. The favorable conditions for the superparaelectricity observation in small ferroelectric nanoparticles

[*] Corresponding author: morozo@i.com.ua; permanent address: V. Lashkarev Institute of Semiconductor Physics, NAS of Ukraine, 41, pr. Nauki, 03028 Kiev, Ukraine

at room temperatures are small Curie-Weiss constants and high nonlinear coefficients as follows from the condition (2). The theoretical forecast is waiting for experimental revealing.

**I. Introduction**

Ferroelectric, ferromagnetic and ferroelastic materials belong to primary ferroics,[1] so that one can expect the similarity of their properties not only in bulk samples, but in nanomaterials also. One of the most interesting and broadly investigated phenomenon in ensemble of ferromagnetic noninteracting nanoparticles was shown to be superparamagnetic phase (see e.g. Ref.[2, 3] and Ref. therein). This phenomenon is related to the fact that for nanoparticle with radius smaller than magnetic exchange length a barrier between different orientations of magnetization is of the order of $k_B T$ at temperatures $T$ < 100 K because it is proportional to the nanoparticle volume. As a result, the particle can be considered as a free reorientable one up to some low enough blocking temperature $T_b$, smaller than the barrier height. At $T < T_b$ magnetic hysteresis loop appears, which is characteristic for monodomain ferromagnetic. Magnetization of the noninteracting nanoparticles in magnetic field has no quantization contrary to paramagnetic molecules. It can be described similarly to classic paramagnetics by Langevin function where the moment of whole particle and not the moment of separate ion has to be considered as elementary magnetic moment.[4] As a result the relaxation time of thermo-activated magnetization is much larger than for conventional paramagnetics.

One could expect the appearance of similar phase in the other primary ferroics, in particular in the ensemble of ferroelectric nanoparticles. Unfortunately up to now nothing is known about superparaelectric phase in ferroelectric nanoparticles, also the term "superparaelectric" was used for description of bulk ferroelectric relaxors.[5], [6] There were the attempts to apply the term to ferroelectric film[7] and to the ferroelectric nanoparticles.[8] However the observed properties were characteristic mainly to paraelectric phase rather than to superparaelectric one.

In this paper we considered for the first time the conditions at which superparaelectric phase would exist in ferroelectric nanoparticles. In our approach we define superparaelectrics characteristic features by analogy with those of superparamagnetics. First of all we took into account that while in magnetic nanopartciles exchange interaction try to align magnetic moments of ions the correlation effect plays the same role in ferroelectric nanopartciles. Actually, polarization fluctuation correlations are determined by the correlation radius[9], originated from the long-range interactions. Therefore in the nanopartcile with radius smaller than correlation radius all the electric dipoles in the particle have to be aligned in the same direction. So we arrive to the pattern of nanoparticles with large enough electric dipoles. The behaviour of the ensemble of such non-interacting nanoparticles under the external electric field, temperature and other factors have to define the size dependent characteristic features of superparaelectric phase. Below we study these features for the case of spherical ferroelectric nanoparticles. In Section II we performed calculations of correlation radius dependence on the particle radius, temperature and ferroelectric material parameters. Potential barrier of polarization reorientation in ferroelectric nanoparticle is calculated in Section III. Polarization dependence on electric field and hysteresis loop calculations allowing for barrier existence is presented in Section IV. Superparaelectricity appearance in ferroelectric nanoparticles is discussed in Section V.

**II. Correlation radius dependence on particle radius**

Let us introduce a correlation function of polarization z-component $P_3(\mathbf{r})$ fluctuations in conventional way [10]

$$G(\mathbf{r},\mathbf{r}') = \langle (P_3(\mathbf{r}) - \langle P_3(\mathbf{r}) \rangle)(P_3(\mathbf{r}') - \langle P_3(\mathbf{r}') \rangle) \rangle, \quad (1)$$

where $\langle ... \rangle$ stands for thermal (statistical) averaging. Using the fluctuation-dissipation theorem, [9, 11] one can rewrite the correlation function (1) via a generalized susceptibility $\chi(\mathbf{r},\mathbf{r}')$ in the form $G(\mathbf{r},\mathbf{r}') = k_B T \chi(\mathbf{r},\mathbf{r}')$, where $\chi(\mathbf{r},\mathbf{r}')$ determines the increment of polarization $\delta P_3(\mathbf{r})$ under the inhomogeneous electric field $\delta E_3(\mathbf{r}')$:

$$\delta P_3(\mathbf{r}) = \int \chi(\mathbf{r},\mathbf{r}') \delta E_3(\mathbf{r}') d\mathbf{r}'. \quad (2)$$

In order to find the generalized susceptibility $\chi(\mathbf{r},\mathbf{r}') \sim G(\mathbf{r},\mathbf{r}')$ of confined system, one has to consider the equation of state for z-component of the polarization $P_3(\mathbf{r}) = P(\mathbf{r}) + \delta P_3(\mathbf{r})$:

$$a_1 P_3 + a_{11} P_3^3 - \zeta \frac{\partial^2 P_3}{\partial z^2} - \eta \left( \frac{\partial^2 P_3}{\partial x^2} + \frac{\partial^2 P_3}{\partial y^2} \right) = E_0 + E_d(P_3) + \delta E_3 \quad (3)$$

Gradient terms $\zeta > 0$ and $\eta > 0$; expansion coefficient $a_{11} > 0$ for the second order phase transitions. Coefficient $a_1(T) = \alpha_T (T - T_c)$, $T_c$ is the transition temperature of bulk material. Note, that the coefficient $a_{11}$ for displacement type ferroelectrics does not depend on $T$, while it is temperature dependent for order-disorder type ferroelectrics [see Appendix A].

$E_0$ is the homogeneous external field, the term $E_d(P_3)$ represents depolarization field, that increases due to the polarization inhomogeneity in confined system. Linear operator $E_d(P_3)$ essentially depends on the system shape and boundary conditions, at that $E_d(0) \equiv 0$. For the most of the cases it has only integral representations, which reduces to constant (depolarization factors) only for special case of ellipsoidal bodies with homogeneous polarization distribution. In this case of polarization dependence on the $x$ and $y$ coordinates simple expression for electric depolarization field, obtained by Kretschmer and Binder[12] is not valid. Below we consider the case when depolarization field is completely screened by the ambient free charges σ outside the particle, while it is nonzero inside the particle due to inhomogeneous polarization distribution [see Fig.1b]. Expression for the spherical particle depolarization field is derived in Appendix B.

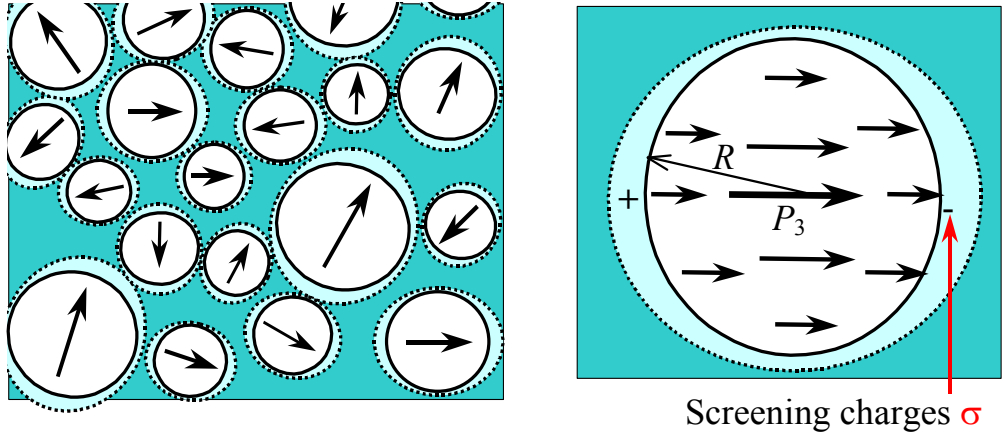

**Figure 1.** (Color online) (a) Ensemble of non-interacting ferroelectric nanoparticles covered outside by the ambient free charges σ. All particle radii $R$ are less than the correlation radius $R_c$, so that the dipole moments inside the particle are aligned due to the correlation effects. (b) A given nanoparticle, where the arrows inside the particle indicate the absolute value of dipole moments in different points.

The boundary conditions depend on the geometry and surface energy of the system depending on the extrapolation length $\lambda$.[18] For the spherical particle of radius $R$ the boundary conditions are:

$$\left(\lambda \frac{dP_3}{dr} + P_3\right)\bigg|_{r=R} = 0 \qquad (4)$$

Where $r = \sqrt{x^2 + y^2 + z^2}$ is radius in spherical coordinates. The typical values are $|\lambda| = 0.5...50$ nm.[13].

Using equation of state (3), one can write the linearized equation for the fluctuation $\delta P_3$ as:

$$\left(a_1 + 3a_{11}P^2(\mathbf{r})\right)\delta P_3 + E_d(\delta P_3) - \zeta\frac{\partial^2 \delta P_3}{\partial z^2} - \eta\left(\frac{\partial^2 \delta P_3}{\partial x^2} + \frac{\partial^2 \delta P_3}{\partial y^2}\right) = \delta E_3. \qquad (5a)$$

The equilibrium polarization $P \equiv P_3^0$ satisfies the nonlinear equation:

$$a_1 P + a_{11} P^3 - \zeta \frac{\partial^2 P}{\partial z^2} - \eta \left( \frac{\partial^2 P}{\partial x^2} + \frac{\partial^2 P}{\partial y^2} \right) = E_0 + E_d(P). \tag{5b}$$

Maxwell's equations $\mathrm{div}(\mathbf{P}(\mathbf{r}) + \varepsilon_0 (\mathbf{E}_d(\mathbf{r}) + \mathbf{E}_0)) = 0$ and $\mathrm{rot}\,\mathbf{E}_d(\mathbf{r}) = 0$ lead to the expression in Fourier **k**-representation: $\delta \widetilde{\mathbf{E}}_d = -\left( \frac{\delta \widetilde{\mathbf{P}}(\mathbf{k}) \mathbf{k}}{\varepsilon_0 \, k} \right) \frac{\mathbf{k}}{k}$ [14]. So, the relation between depolarization field fluctuations $\widetilde{E}_d$ and polarization $\delta P_3(\mathbf{r})$ is:

$$\widetilde{E}_d = -\frac{k_3^2}{\varepsilon_0 k^2} \delta \widetilde{P}_3(\mathbf{k}). \tag{6}$$

In Appendix B2 we had shown the validity of Eq.(6) for nanoparticles.

Using the definition of generalized susceptibility (2) and Eqs.(5)-(6), we obtained the approximate solution for linearized susceptibility $\widetilde{\chi}(\mathbf{k}) \sim \widetilde{G}(\mathbf{k})$ in Fourier **k**-representation as:

$$\widetilde{\chi}(\mathbf{k}) \approx \frac{1}{3 a_{11} \overline{P}^2 + a_R(T, R) + \eta(k_1^2 + k_2^2) + (\zeta + 1/\varepsilon_0 k^2) k_3^2}. \tag{7}$$

Eq.(7) works well for low **k** values (long-wave approximation). $\overline{P}$ is the equilibrium polarization averaged over the nanoparticle volume satisfies the equation $a_R(T, R) \overline{P} + a_{11} \overline{P}^3 = E_0$. At zero external field $E_0 = 0$, the spontaneous polarization is nonzero in ferroelectric phase, $\overline{P}^2 = -a_R(T, R)/a_{11}$, while $\overline{P} = 0$ in paraelectric phase. Depolarization field and surface influence lead to the renormalization of coefficient $a_1$ in Eqs.(7) as[15]

$$a_1 \to a_R(T, R) \approx \alpha_T \left( T - T_c \left( 1 - \frac{R_{cr}^0}{R} \right) \right). \tag{8}$$

Where $R_{cr}^0 = \dfrac{3\xi}{\alpha_T T_c \left( \lambda + \sqrt{\xi \varepsilon_0} \right)}$ is a critical radius of size-induced paraelectric phase appearance at zero temperature and $R < R_{cr}$. It exists at $\lambda > -\sqrt{\xi \varepsilon_0}$, and thus we consider the case of the positive extrapolation length hereinafter. At a given temperature $T$ the sphere critical radius $R_{cr}(T) \approx \dfrac{R_{cr}^0}{1 - T/T_c}$ exists at $T < T_c$ and should be found from the condition $a_R(T, R_{cr}) = 0$. Both measured and calculated values of $R_{cr}$ typically depend on temperature $T$ and varied within the range 2-50 nm [16,17][18]. At radiuses $R >> R_{cr}(T)$ the particles ferroelectric properties are close to the bulk material.

In plane, perpendicular to the polar axis ($k_3 = 0$) the expression (7) can be simplified to a Lorenzian form $\tilde{\chi}(\mathbf{k}) \sim \left(R_{cx}^2(k_1^2 + k_2^2) + 1\right)^{-1}$, where the correlation radii for the fluctuations across polar axis are introduced as:

$$R_{cx}(T,R) = \begin{cases} \sqrt{\dfrac{\eta}{a_R(T,R)}}, & a_R(T,R) > 0, \\ \sqrt{\dfrac{-\eta}{2a_R(T,R)}}, & a_R(T,R) < 0. \end{cases} \quad (9)$$

Since usually gradient coefficients $\xi \sim \eta$, while for ferroelectrics $a_1\varepsilon_0 \ll 1$, the longitudinal correlation radius $R_{cz}$ is almost constant and much smaller than $R_{cx}$. This means that the depolarization field suppresses longitudinal fluctuations.[14]

Transverse correlation radius $R_{cx}$ dependence via the particle radius calculated on the basis of Eq.(9) is shown in Fig.2. At temperatures $T<T_c$ transverse correlation radius $R_{cx}$ diverges at critical radius $R_{cr}(T)$ as anticipated from Eq.(9) (see curves 1-3). The divergence corresponds to the size-induced ferroelectric-paraelectric phase transition. At temperatures $T>T_c$ transverse correlation radius monotonically increases with the particle radius because of $a_R^{-1}(T,R)$ increase (see curves 4-5).

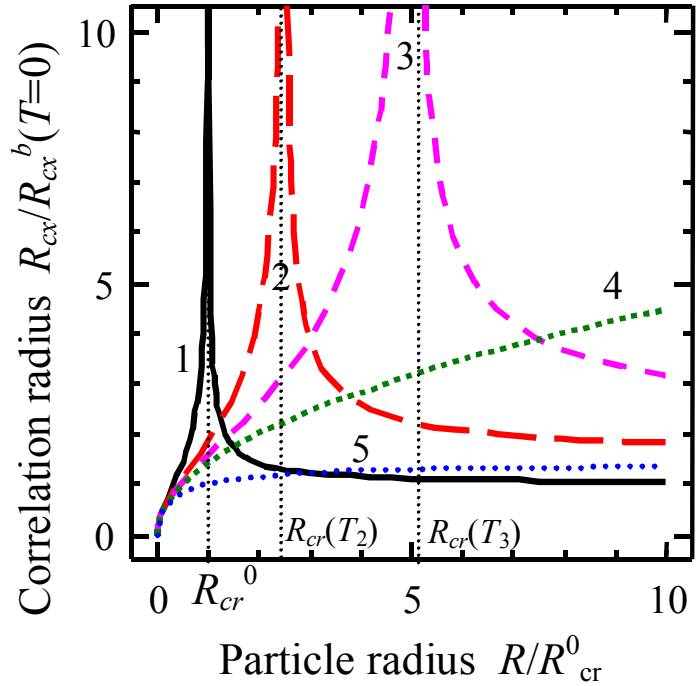

**Figure 2.** (Color online) Transverse correlation radius $R_{cx}/R_{cx}^b(T=0)$ dependence on particle radius $R/R_{cr}^0$ for different temperatures below ($T<T_c$), equal ($T=T_c$) and above ($T>T_c$) the bulk

transition temperature $T_c$: $T/T_c$=0, 0.6, 0.8, 1, 2 (curves 1, 2, 3, 4, 5). Bulk correlation radius $R_{cx}^b = \sqrt{-\eta/2a_1(T)}$ is defined in ferroelectric phase ($T<T_c$).

**III. Potential barrier for polarization reorientation in ferroelectric nanoparticles**

In the vicinity of the phase transition polarization reorientation is caused by fluctuations. So, let us study the question how high can be potential barrier of polarization reorientation in spherical ferroelectric nanoparticles under the absence of external field.

*III.1. Barrier for perovskite and uniaxial ferroelectric nanoparticles*

For uniaxial ferroelectrics the barrier between the states $\pm P_0$ can be estimated on the basis of the free energy:

$$F(\overline{\mathbf{P}}) = V\left(\frac{a_R(T,R)}{2}\overline{P}^2 + \frac{a_{11}}{4}\overline{P}^4 - \overline{P}E_0\right). \tag{10}$$

In Eq.(10) the integration over the spherical particle volume $V = 4\pi R^3/3$ was performed. At zero external electric field ($E_0 = 0$) the barrier $\Delta F = V a_R^2(T,R)/4a_{11}$.

For perovskite ferroelectrics the orientation barrier can be estimated on the basis of the free energy:

$$F(\overline{\mathbf{P}}) = V\left(\begin{array}{c}\dfrac{a_R(T,R)}{2}\left(\overline{P}_1^2 + \overline{P}_2^2 + \overline{P}_3^2\right) + \dfrac{a_{12}}{2}\left(\overline{P}_1^2\overline{P}_2^2 + \overline{P}_3^2\overline{P}_2^2 + \overline{P}_3^2\overline{P}_1^2\right) + \\ + \dfrac{a_{11}(T,R)}{4}\left(\overline{P}_1^4 + \overline{P}_2^4 + \overline{P}_3^4\right) - \overline{P}_1 E_{01} - \overline{P}_2 E_{02} - \overline{P}_3 E_{03}\end{array}\right) \tag{11}$$

Free energy (11) is stable only for $a_{11} > 0$ and $a_{11} + 2a_{12} > 0$, otherwise one should consider higher order terms in Eq. (11). At zero external electric field, $E_0 = 0$, free energy (11) can describe paraelectric phase (PE) with $\overline{P}_i = 0$ (thermodynamically stable at $a_R > 0$); rhombohedral ferroelectric phase (rFE) with $\overline{P}_1^2 = \overline{P}_2^2 = \overline{P}_3^2 = -a_R/(a_{11} + 2a_{12})$ (stable at $a_R < 0$ and $a_{11} > a_{12}$); tetragonal ferroelectric phase (tFE) with $\overline{P}_i^2 = -a_R/a_{11}$, $\overline{P}_j^2 = \overline{P}_k^2 = 0$, $i \neq j \neq k$ (stable at $a_R < 0$ and $a_{11} < a_{12}$). The saddle points $\overline{P}_i^2 = \overline{P}_j^2 = -a_R/(a_{11} + a_{12})$, $\overline{P}_k^2 = 0$, bordering the minima exist in tFE. One can find the potential barrier between different polarization orientations in stable phases as a difference between the free energy values corresponding to minimum and saddle point. Thus, using Eqs (10)-(11), we have found the reorientation barrier in the form:

$$\Delta F(T,R) = \gamma(T,R) \cdot k_B T,$$

$$\gamma(T,R) = V \frac{a_R^2(R,T)}{4 k_B T \cdot a_{11}} \begin{cases} \dfrac{a_{12} - 2a_{11}}{a_{12} + 2a_{11}}, & 2a_{11} < a_{12}, \quad \text{perovskite FE}, \\ \dfrac{a_{11}}{a_{11} + a_{12}} \left( \dfrac{2a_{11} - a_{12}}{a_{12} + 2a_{11}} \right), & 2a_{11} > a_{12}, \quad \text{perovskite FE}, \\ 1, & \text{uniaxial FE}. \end{cases} \qquad (12)$$

Here the dimensionless barrier height $\gamma$ is introduced.

Note, that Eq. (11) transforms into Eq. (10) at $a_{11} = a_{12}$ with accuracy of multiplayer 3. In this "isotropic" case the barrier and the circle of minima looks like "sombrero".

Examples of the phase diagrams in coordinates temperature – size are shown in Fig.3 for perovskite (a,b) and uniaxial (c,d) ferroelectric nanopaticles. Solid curves correspond to the transition between paraelectric and ferroelectric phases, i.e. it is the dependence $T_{cr}(R)$ and so $\Delta F(T_{cr}, R)=0$. Dashed curves represent the situation when the reorientation barrier is equal to the energy of thermal fluctuations, $\Delta F=k_B T$; dotted curves correspond to barriers $\Delta F=2k_B T$ (a,b) and $50k_B T$ (c,d). Filled area between corresponding solid and dashed curves indicates the regions with potential barrier $\Delta F$ lower that the thermal activation energy.

Freezing temperature $T_f(R_0)$ at a given particle radius $R_0$ can be as estimated an intersection of the vertical line $R=R_0$ with corresponding dashed (or dotted) curves as shown in Figs.3a. So, at a given particle radius $R_0$ **free reorientation** of polarization is expected in the temperature range $T_f(R_0)<T<T_{cr}(R_0)$. Similarly, at fixed temperature $T_0$ the "freezing" nanoparticle radius $R_f(T_0)$ can be estimated an intersection of the horizontal line $T=T_0$ with corresponding dashed (or dotted) curves shown in Figs.3a. Surely the range determination is somehow voluntary, since the value of $T_f(R)$ or $R_f(T)$ depends on numerical factor $\gamma$ before $k_B T$ that in turn depends on the system characteristics.

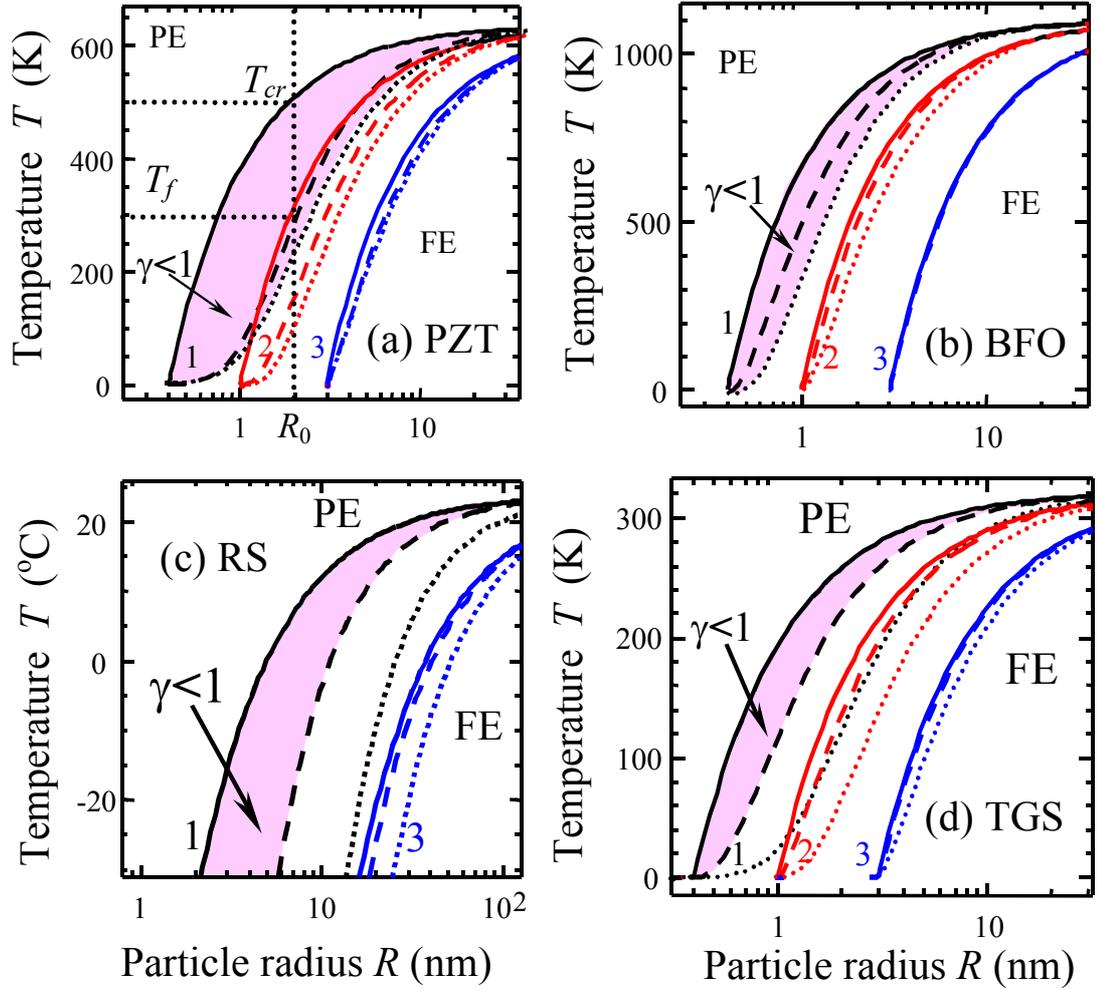

**Figure.3.** (Color online) Phase diagram in coordinates temperature – size for four different set of material parameters corresponding to PbZr$_{0.6}$Ti$_{0.4}$O$_3$ (PZT), BiFeO$_3$ (BFO), Rochelle Salt (RS) and triglycine sulfate (TGS) (panels a, b, c and d respectively). Solid curves are related to the transition between paraelectric (PE) and ferroelectric (FE) phases, i.e. it is $T_{cr}(R)$, where $\Delta F=0$. Dashed curves represent the situation when the reorientation barrier is equal to the energy of thermal fluctuations, $\Delta F=k_BT$; dotted curves correspond to barriers $\Delta F=2k_BT$ (a,b) and $50k_BT$ (c,d). Groups of curves 1, 2 and 3 correspond to critical radius at zero temperature $R_{cr}^0 =0.4$, 1 and 3 nm. Coefficients for BiFeO$_3$ were taken from Ref. [19], namely $a_1 = 9.8(T-1103)\cdot 10^5$ ($C^{-2}m^2N$) with temperature in Kelvin, $a_{11}=13\cdot 10^8$ ($C^{-4}m^6N$), $a_{12}=2\cdot 10^8$ ($C^{-4}m^6N$). Coefficients for PbZr$_x$Ti$_{1-x}$O$_3$ can be found in Ref.[20]. Free energy expansion coefficients for RS and TGS were calculated from data given in Ref. [21].

So at fixed radius $R$, superparaelectric phase (**SPE**) may appear only in the temperature range $T_f(R)<T<T_{cr}(R)$. At fixed temperature (e.g. at room) **SPE** may appear only at nanoparticle radiuses $R_{cr}(T)<R<R_f(T)$. However, as was stated in the Introduction, the condition of all ions inside the particle alignment is necessary for SPE appearance [see Fig.1b]. As we have discussed in the Introduction to superparamagnetic nanoparticles, the alignment due to the correlation effects is possible, when the particle radius $R$ is less than the correlation radius $R_c$. On the other hand the particle radius must be higher than the critical radius $R_{cr}$ of size-driven ferroelectric-paraelectric phase transition. Therefore we have to find out the region of radii where the conditions can be fulfilled.

### III.2. Correlation effect in ferroelectric nanoparticles

The dependence of ratio $R_{cx}(T,R)/R$ on particle radius $R$ for the temperatures $T$ corresponding to orientation barrier $\Delta F(T,R) = k_B T$ (solid curves) and $\Delta F(T,R) = 10 k_B T$ (dotted curves) are shown in Figs.4a,b for PZT and TGS respectively. In the range $\infty > R_{cx}(T,R)/R > 1$ the alignment of elementary dipoles appears due to the correlation effects as well as adopted here long-wave approximation is valid and self-consistent. So, the filled area between solid curves ($\Delta F(T,R) = k_B T$), dashed vertical lines of size-induced FE-PE phase transition ($R = R_{cr}^0$ and $\Delta F=0$) and horizontal line ($R_{cx}=R$) indicates the region of particle radii corresponding to the possible superparaelectric phase appearance.

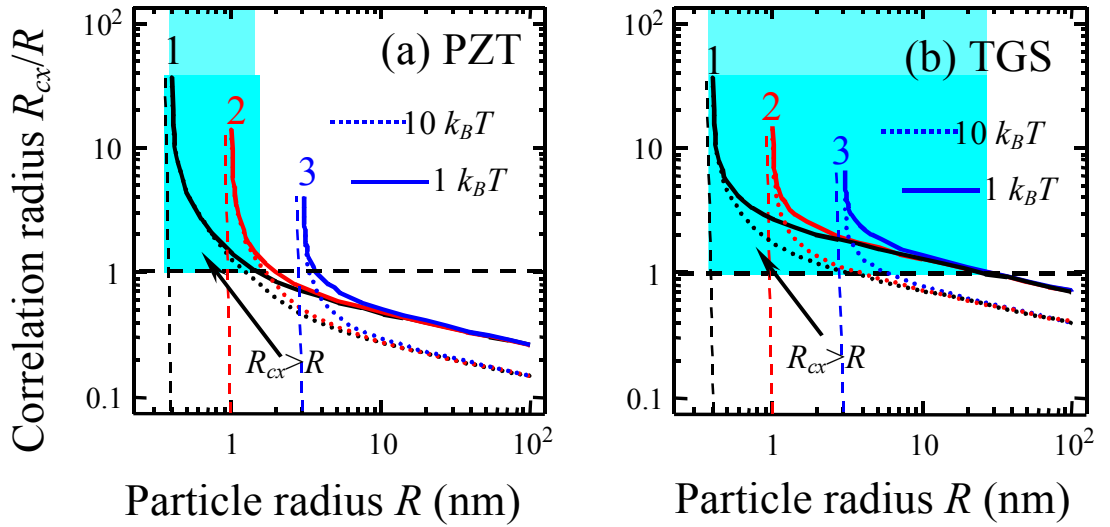

**Figure 4.** (Color online) (a,b) The dependence of ratio $R_{cx}(T,R)/R$ on particle radius $R$ for orientation barrier $k_B T$ (solid curves) and $10 k_B T$ (dotted curves) and different values of critical radius $R_{cr}^0$ =0.4, 1 and 3 nm (curves 1, 2, 3). Vertical lines indicate the FE-PE phase transition appearing at critical radius $R_{cr}^0$; $R_{cx}^b(T=0)$=1 nm (other material parameters are the same as in Fig.3). Filled area reflects the region of superparaelectric phase existence.

It is seen from Fig.4 that the region of SPE phase can be large enough, its width being essentially dependent on existence of low barrier; and the region is broader for TGS than for PZT.

In agreement with Eq.(12), energetic barrier height $\Delta F(R) \sim R^3 \alpha_T^2/a_{11} \sim V \cdot R_{cx}^{-2} \overline{P}_0^2$. So to obtain low barriers in the wider vicinity of PE-FE size-induced phase transition, high $R_{cx}$, low $\overline{P}_0^2$ values and small enough particle radii are necessary. The favorable conditions for small $\Delta F$ and high $R_{cx}$ values are small Curie-Weiss constants and high coefficients $a_{11}$.

To summarize, the condition $0 < \Delta F(T,R) \leq k_B T$ determines the superparaelectric phase appearance in ferroelectric nanoparticles of radius $R_{cr}(T) < R < R_f(T) < R_c(T)$ at given temperature $T$; or alternatively at temperatures $T_f(R) < T < T_{cr}(R)$ at given radius $R$. In the region of relatively low potential barrier of polarization reorientation under *external electric field* could partially or fully align the nanoparticles dipoles as considered below.

We have to underline that namely aforementioned temperature – size region corresponds to the true SPE phase. In this region polarization of free orientable dipoles has to be independent on the regime of external electric field application (field cooling, zero field cooling, etc) so here the behavior of SPE is ergodic. In what follows we will consider the influence of external electric field on polarization for two cases, namely for $k_B T$ larger and smaller than the barrier for dipole reorientation. In the latter case the hysteresis loop can appear at $T < T_f(R)$ that is known to be characteristic feature of ferroelectric phase. One can also expect the transformation of behavior into nonergodic one with dependence of polarization response on the regime of field application like it was observed for superparamagnetic phase.[22] On the other hand the behavior in the region $T < T_b$ is considered as "blocked superparamagnetizm" in Ref.[4]. In any case it is obvious that the region $T < T_f$ has to be included into consideration of superparaelectrics.

**IV. Polarization response to external field**

At a given temperature $T$ and nanoparticle radius $R$, polarization orientation with respect to external field $\mathbf{E}_0 = \{0, 0, E_0\}$ can be written as $\mathbf{P} = \{P\sin\theta\cos\varphi, P\sin\theta\sin\varphi, P\cos\theta\}$. After elementary transformations the free energy (10)-(11) acquires the form:

$$F(\overline{P},\theta,\varphi) = V \begin{cases} \left(\dfrac{a_R(T,R)}{2}\overline{P}^2 + \dfrac{a_{11}}{4}\overline{P}^4 - E_0\overline{P}\cos\theta + \right. \\ \left. \dfrac{(a_{12}-a_{11})}{2}\overline{P}^4\left(\sin^4\theta\sin^2\varphi\cos^2\varphi + \sin^2\theta\cos^2\theta\right)+\right), & \text{perovskite FE,} \\ \left(\dfrac{a_R(T,R)}{2}\overline{P}^2 + \dfrac{a_{11}}{4}\overline{P}^4 - E_0\overline{P}\cos\theta\right), & \text{uniaxial FE.} \end{cases} \quad (13)$$

Here $\{\theta, \varphi\}$ are the spherical angles. Similarly to the consideration of superparamagnetics proposed by Binder and Young,[3] in order to obtain _equilibrium_ polarization vector $\langle \mathbf{P}(E_0) \rangle = \{P_1, P_2, P_3\}$, polarization components $P_i(E_0, \theta, \varphi)$ should be averaged over the spherical angles $\{\theta, \varphi\}$ and polarization absolute value $\overline{P}$ distribution:

$$\langle \overline{P_i}(E_0) \rangle = \int_0^\infty d\overline{P} \cdot \mu(\overline{P}) \exp\left(-\frac{F_0(\overline{P})}{k_B T}\right) \frac{\int_0^{2\pi} d\varphi \int_0^\pi d\theta \sin\theta \exp(-F_a(\overline{P},\theta,\varphi)/k_B T) \cdot \overline{P_i}(E_0,\theta,\varphi)}{\int_0^{2\pi} d\varphi \int_0^\pi d\theta \sin\theta \cdot \exp(-F_a(\overline{P},\theta,\varphi)/k_B T)}. \quad (14)$$

Here $\mu(\overline{P})$ is the distribution function of $\overline{P}$, originated from the dispersion of nanoparticle sizes $R$, schematically shown in Fig.1a.

$F_0(\overline{P}) = V(R)\left(\dfrac{a_R(R,T)}{2}\overline{P}^2 + \dfrac{a_{11}}{4}\overline{P}^4\right)$ is the isotropic part (i.e. angles-independent) of the free energy (13) that coincides with Eq.(10) at $E_0=0$; $F_a(\overline{P},\theta,\varphi)$ is the angle-dependent part. Note, that due to the conservation of magnetization vector absolute value, the isotropic part $F_0$ does not play any important role in superparamagnetics consideration (para-process is typically small) [1-2]. However, in ferroelectrics polarization absolute value strongly depends on the boundary conditions and applied electric field, so $F_0$ is essential.

The averaging over the spherical angles in Eq.(14) leads to the dependences $\langle P_1(E_0) \rangle = \langle P_2(E_0) \rangle = 0$.

Note, that _equilibrium_ (or _absolutely stable_) remnant polarization is always absent, i.e. $\langle P_3(E_0 = 0) \rangle = 0$. Actually, the free energy (13) is the even function of $\cos\theta$ at $E_0 = 0$, while $P_3 = \overline{P}\cos\theta$ is an odd function of $\cos\theta$, and so at zero applied field only the trivial solution $\langle P_3(E_0 = 0) \rangle = \overline{P}\langle \cos\theta \rangle \equiv 0$ exists. The mathematical result reflects the conventional statement that _equilibrium_ polarization of perfect ferroelectrics can be reversed by infinitely small field, applied long enough.[21], [23] In other words, hysteresis phenomenon appearence at $T<<T_f(R)$ (or nanoparticle radiuses $R>>R_f(T)$) has to correspond to the _metastable (non-ergodic) state_. In order to include bistable states (e.g. hysteresis) averaged values $\langle P_i \rangle$ should be substituted either into Eq.(14) or in the free energy (13) *self-consistently* [10].

In what follows we will consider separately the nanoparticles of the uniaxial and perovskite ferroelectrics allowing for the different form of their free energy and so barriers.

*IV.1. Uniaxial ferroelectrics*

For uniaxial ferroelectrics, analytical integration on the spherical angles $\{\theta, \varphi\}$ can be performed in Eq.(14). This leads to the expression:

$$\langle \overline{P}_3(E_0) \rangle = \int_0^\infty d\overline{P} \cdot \mu(\overline{P}) \overline{P} \exp\left(-\frac{F_0(\overline{P})}{k_B T}\right) \cdot L\left(\frac{VE_0 \overline{P}}{k_B T}\right),$$

$$L(x) = \frac{1}{\tanh(x)} - \frac{1}{x}.$$
(15)

Function $L(x)$ is exactly Langevin function. The independent on electric field and related to the barrier exponential term with isotropic part of the free energy $F_0$ is the additional factor in comparison with superparamagnetics.

It is clear from Eq.(15) that the Langevin law can be obtained for infinitely-sharp Dirac-delta distribution function $\mu(P) = \delta(P - \overline{P})$ (i.e. all particles have equal radius) and low energetic barriers $|F_0(\overline{P})| \ll k_B T$ so $\exp(-F_0/k_B T) \approx 1$ (i.e. particle radius is much smaller than the freezing radius $R \ll R_f(T)$), namely for the case:

$$\langle \overline{P}_3(E_0) \rangle \approx \overline{P} \tanh^{-1}\left(\frac{VE_0 \overline{P}}{k_B T}\right) - \frac{k_B T}{VE_0}.$$
(16)

Surely, in a given nanocomposite sample nanoparticle sizes are distributed. For the case the averaging in Eq.(15) with finite-width distribution function $\mu(P)$ should be performed. Obtained results essentially depend on distribution function parameters, however in any case one can obtain Langevin-like behavior for sharp enough distribution function $\mu(P)$ and low energetic barriers. Langevin-like behavior in superparamagnetic was obtained in Ref.[4]; the phenomenon was called isotropic supermagnetizm.

At fixed temperature (e.g. at room) and nanoparticle radius essentially higher than the "freezing radius" $R \gg R_f(T)$, orientation barrier is much higher that the thermal fluctuations $k_B T$. So, lets divide the integration on the particle radius in Eq.(15) into two regions $0 < R < R_f(T)$ and $R > R_f(T)$. The region $0 < R < R_f(T)$ (and so $R < R_c$ since $R_f < R_c$) corresponds to the behavior without hysteresis, since here the barrier $|F_0(\overline{P})| < k_B T$ and so exp~1. The region $R \gg R_f(T)$ (and so $R \gg R_c$), where $|F_0(\overline{P})| \gg k_B T$, has to be ferroelectric phase that could be considered self-consistently, namely with the substitution $P^4 \sim P^2 \langle P_3 \rangle^2$ in the free energy $F_0$. Using these ideas and Laplace method of integration, we obtained approximate analytical expressions:

$$\left\langle \overline{P}_3(E_0) \right\rangle \approx \int_0^{R_f(T)} dR \cdot \tilde{\mu}(R) \overline{P}(R) L\left( \frac{VE_0 \overline{P}(R)}{k_B T} \right) + \int_{R_f(T)}^{\infty} \frac{dR \cdot \tilde{\mu}(R) \cdot E_0}{a_R(R,T) + a_{11} \left\langle \overline{P}_3(E_0) \right\rangle^2}. \qquad (17)$$

Here $\tilde{\mu}(R)$ is the normalized distribution function of nanoparticle radii $R$, related with $\mu(\overline{P})$ in conventional way (see e.g. Refs. [24], [25]), $\overline{P}$ satisfies the equation $a_R(T,R)\overline{P} + a_{11}\overline{P}^3 = E_0$. Note, that $R_{cr}(T) < R_f(T) < R_c(T)$ as stated in Section III. Thus in the first integral $R < R_c$, while in the second integral both regions $R < R_c$ and $R > R_c$ are included.

Dependence of mean polarization $\langle P_3 \rangle$ on the applied electric field in shown in Figs.5a for uniaxial RS material parameters, Dirac-delta distribution $\tilde{\mu}(R) = \delta(R - R_0)$ of particle sizes, fixed freezing radius $R_f(T)$ at room temperature $T=0^\circ C$, and different nanoparticle average radius $R_0$. Curves 1-4 for $R_0 < R_f$ correspond to the Langevin law, while the curves 5-7 for $R_0 > R_f$ indicate the hysteresis loop appearance.

Solid curves in Fig.5b are Langevin law $\left\langle \overline{P}_3(E_0) \right\rangle$ for different nanoparticle radius $R$; dashed curve corresponds to the rectangular distribution function $\tilde{\mu}(R) = 1/R_f$ at $0 < R < R_f$, shown in the inset.

Dependence $\left\langle P_3(E_0) \right\rangle$ calculated from Eq.(17) for bell-shaped distribution functions $\tilde{\mu}(R)$ is shown in Fig.5c (curves 1-3). Dotted curves correspond to Dirac-delta distribution $\tilde{\mu}(R) = \delta(R - R_0)$ for small $R_0 < R_f$ and high $R_0 > R_f$.

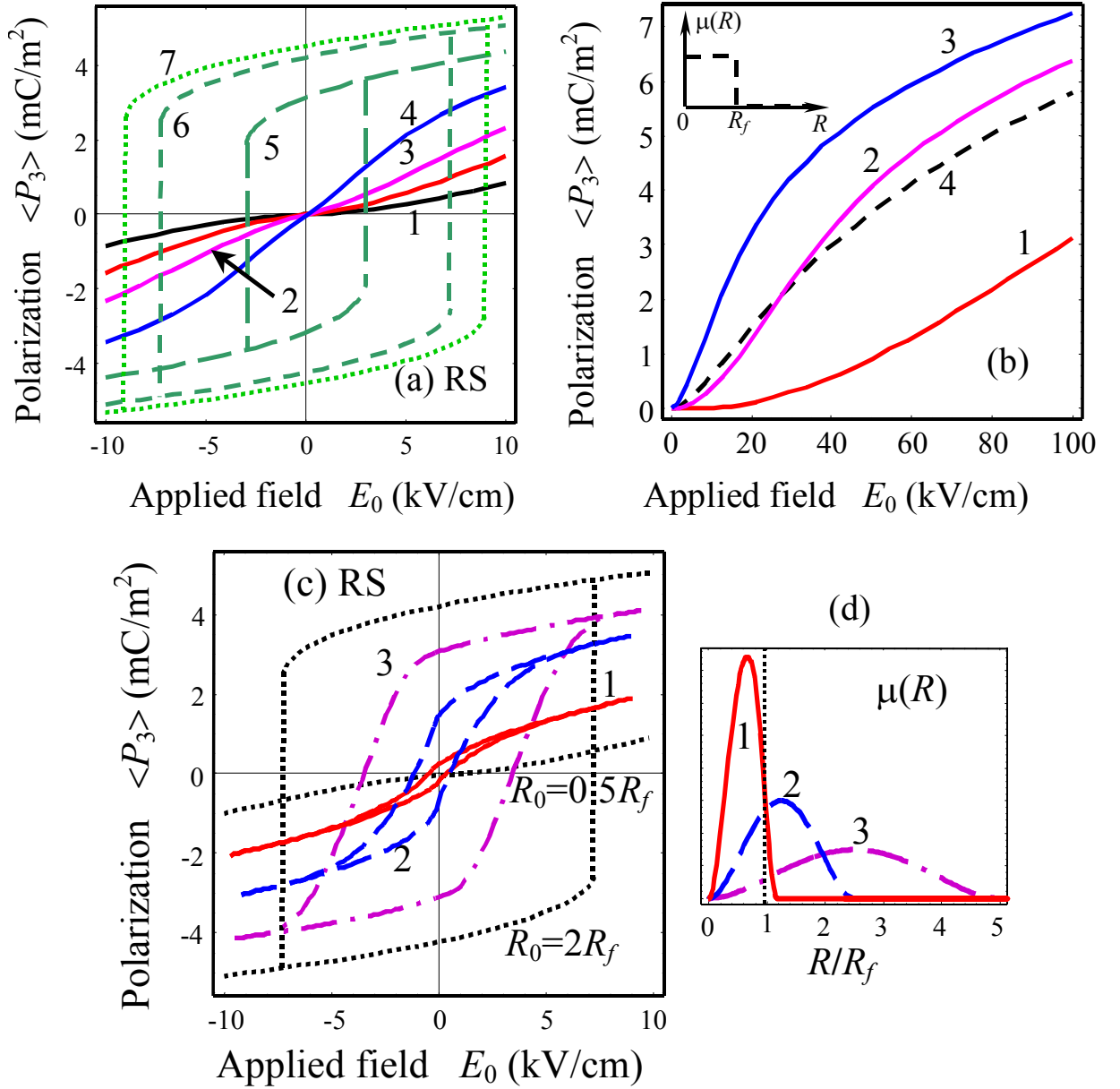

**Figure 5.** (Color online) (a) Dependence of Rochelle Salt (RS) polarization $\langle P_3(E_0)\rangle$ on the applied electric field $E_0$ calculated from Eqs. (17) for Dirac-delta distribution $\tilde{\mu}(R)=\delta(R-R_0)$ and different values $R_0$=6, 7, 8, 9, 11, 20, 30 nm (figures 1-7 near curves). The freezing radius $R_f$=10 nm at temperature $T$=273 K, $R_{cr}^0$=0.5 nm (b) Dependence $\langle P_3(E_0)\rangle$: solid curves 1-3 are Langevin law (16) for different nanoparticle radius $R$=3, 5, 7 nm; dashed curve 4 corresponds to Eq.(17) with rectangular distribution function, $\tilde{\mu}(R)=1/R_f$ at $0<R<R_f$, shown in the inset. (c) Dependence $\langle P_3(E_0)\rangle$ calculated from Eq.(17) for well-localized distribution functions $\tilde{\mu}(R)$ (curves 1-3) shown in inset (d). Dotted curves correspond to Dirac-delta distribution $\tilde{\mu}(R)=\delta(R-R_0)$ for $R_0$=0.5$R_f$ and $R_0$=2$R_f$ (labels near the curves). Other material parameters are the same as in Fig.3c.

It follows from Figs.5 that for Langevin-like behavior besides small enough nanoparticles radii and barrier (as it was discussed earlier) the narrow distribution function of the particles radii and so polarization is necessary, the particular form of distribution function being not especially important. It is obvious also, that hysteresis loop originates from contribution of either large enough average particles radii (see Fig.5a) or broad distribution function (see Figs.5c,d). The important role of large particles in appearance of hysteresis loop is similar to what obtained in superparamagnetism earlier [4]. Allowing for this behavior at $T < T_b$ was named as blocked superparamagnetism, it seems reasonable to name the region $T < T_f$ with hysteresis loop as frozen superparaelectricity.

*IV.2. Perovskite ferroelectrics*

For perovskite ferroelectrics with $a_{11} \neq a_{12}$, the averaging on the angle $\varphi$ in Eq.(14) leads to the expression:

$$\langle \overline{P}_3(E_0) \rangle = \int_0^\infty d\overline{P} \cdot \mu(\overline{P}) \exp\left(-\frac{F_0(\overline{P})}{k_B T}\right) \overline{P} \frac{\int_0^\pi d\theta \sin\theta \cos\theta \cdot K(\overline{P},\theta)}{\int_0^\pi d\theta \sin\theta \cdot K(\overline{P},\theta)},$$

$$K(\overline{P},\theta) = 2\pi \exp\left(-\frac{\Delta F_a(\overline{P},\theta)}{k_B T}\right) I_0\left(V \frac{(a_{12}-a_{11})\overline{P}^4}{16 k_B T} \sin^4\theta\right), \quad (18)$$

$$\Delta F_a(\overline{P},\theta) = V\left(\frac{a_{12}-a_{11}}{2}\overline{P}^4 \sin^2\theta \cos^2\theta + \frac{(a_{12}-a_{11})\overline{P}^4}{16}\sin^4\theta - E_0 \overline{P}\cos\theta\right).$$

$I_0$ is the modified Bessel function of zero order.

Similarly to the case of uniaxial ferroelectrics, lets divide the averaging on the particle radius in Eq.(18) into two regions $0<R<R_f(T)$ and $R>R_f(T)$. The region $0<R<R_f(T)$ the barrier $|F_0(\overline{P})| < k_B T$ and so exp~1. The ferroelectric region $R>>R_f(T)$, where $|F_0(\overline{P})| >> k_B T$, was considered self-consistently. Similarly to Eq.(17), we obtained approximate analytical expressions:

$$\langle \overline{P}_3(E_0) \rangle \approx \int_0^{R_f(T)} dR \cdot \widetilde{\mu}(R) \overline{P} \frac{\int_0^\pi d\theta \sin\theta \cos\theta \cdot K(\overline{P},\theta)}{\int_0^\pi d\theta \sin\theta \cdot K(\overline{P},\theta)} + \int_{R_f(T)}^\infty \frac{dR \cdot \widetilde{\mu}(R) \cdot E_0}{a_R(R,T) + a_{11}\langle \overline{P}_3(E_0)\rangle^2}. \quad (19)$$

Here $\widetilde{\mu}(R)$ is the normalized distribution function of nanoparticle radii $R$, $\overline{P}$ satisfies the equation $a_R(T,R)\overline{P} + a_{11}\overline{P}^3 = E_0$.

Dependence of mean polarization $\langle P_3 \rangle$ on the applied electric field is shown in Figs.6a for perovskite PZT material parameters, Dirac-delta distribution $\widetilde{\mu}(R) = \delta(R - R_0)$, fixed temperature $T<T_c$, freezing radius $R_f(T)$ and different nanoparticle radius $R_0$. Curves 1-4 for $R_0<R_f$ correspond to the Langevin-like law, while the curves 5-7 for $R_0>R_f$ indicate the hysteresis loop appearance. Solid curves in Fig.6b are Langevin-like law $\langle P_3(E_0) \rangle$ for different nanoparticle radius $R$; dashed curve corresponds to the rectangular distribution function $\widetilde{\mu}(R)$ shown in the inset. Dependence $\langle P_3(E_0) \rangle$ calculated from Eq.(19) for well-localized distribution functions $\widetilde{\mu}(R)$ is shown in Fig.5c (curves 1-3).

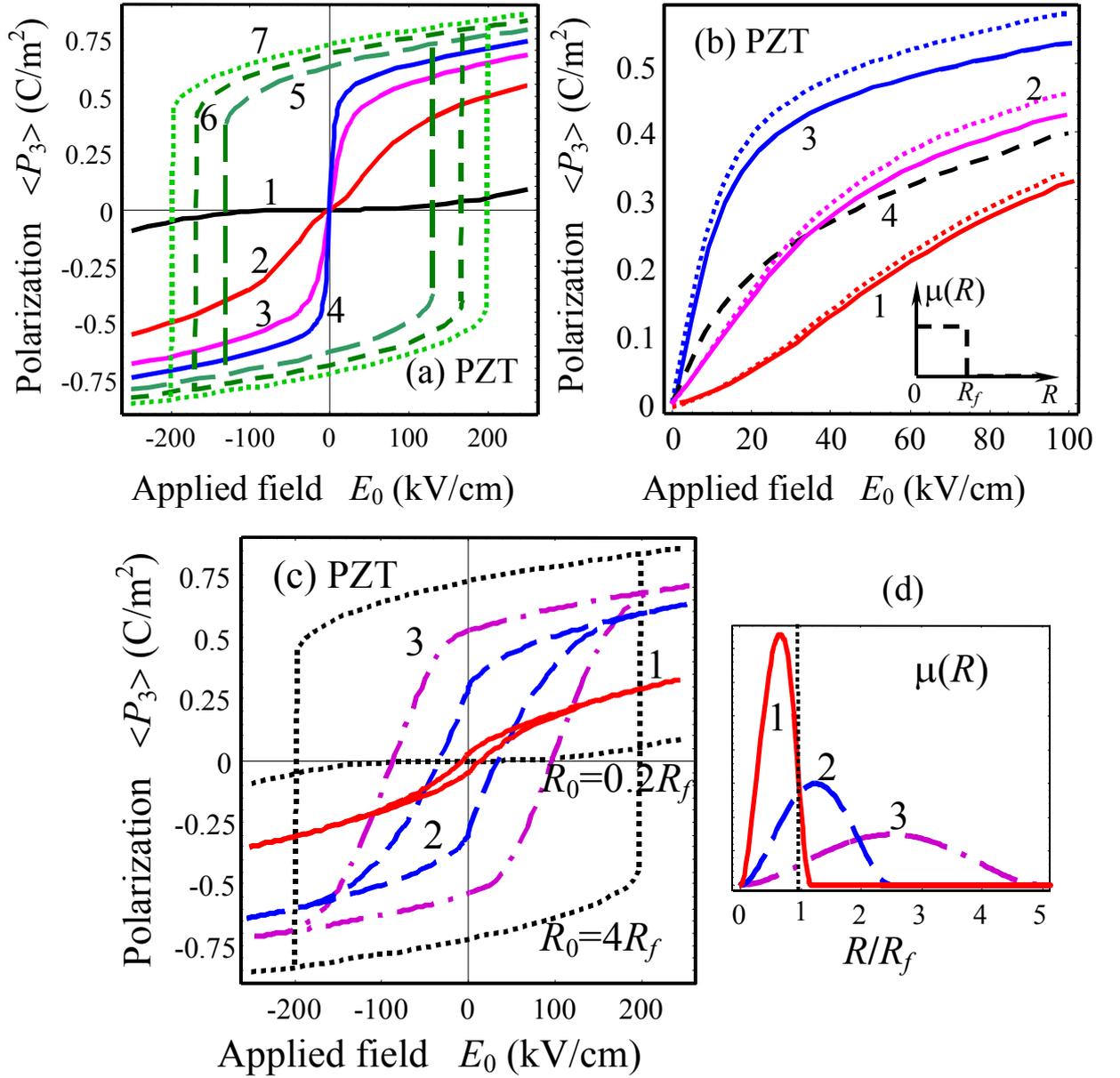

**Figure 6.** (Color online) (a) Dependence of PbZr$_{0.6}$Ti$_{0.4}$O$_3$ (PZT) polarization $\langle P_3(E_0) \rangle$ on the applied electric field $E_0$ calculated from Eqs. (19) for Dirac-delta distribution $\tilde{\mu}(R) = \delta(R - R_0)$ and different values $R_0$=0.5, 1, 1.5, 2, 3, 5, 10 nm (figures 1-7 near curves). The freezing radius $R_f$=2.5 nm at room temperature $T$=293 K, $R_{cr}^0$=0.5 nm. (b) Dependence $\langle P_3(E_0) \rangle$: curves 1-3 are Langevin law (doted curves) and Langevin-like law (solid curves) calculated from Eq.(19) for different nanoparticle radius $R$=1, 1.2, 1.6 nm; dashed curve 4 corresponds to the rectangular distribution function, $\tilde{\mu}(R) = 1/R_f$ at $0<R<R_f$, shown in the inset. (c) Dependence $\langle P_3(E_0) \rangle$ calculated from Eq.(19) for well-localized distribution functions $\tilde{\mu}(R)$ (curves 1-3) shown in inset (d). Dotted curves correspond to Dirac-delta distribution $\tilde{\mu}(R) = \delta(R - R_0)$ for $R_0$=0.2$R_f$ and $R_0$=4$R_f$ (labels near the curves). Other material parameters are the same as in Fig.3a.

It follows from Fig.6 that qualitatively the curves look like those depicted in Fig.5 for uniaxial case. Namely, the narrower the distribution function of sizes the better is the condition of Langevin type behavior observation.

One can see from the Fig.6b that pure Langevin curves are higher than solid curves for different radii. Similarly to superparamagnetics case [4], this behavior could be named as *anisotropic superparaelectricity* because of barrier anisotropy ($a_{11}$-$a_{12}$) contribution in perovskites (see Eq.(18)).

To summarize the Section IV, Langevin-like law for polarization dependence on external field was predicted in the temperature range $T_{cr}(R)>T>T_f(R)$ at fixed radius $R<R_f$ (or in the range $R_{cr}(T)<R<R_f(T)$ at fixed temperature $T$). Bistable remnant polarization appeared at temperatures $T<T_f(R)$ (or at radii $R>>R_f(T)$). For nanoparticle radius less than critical both hysteresis and its precursor, Langevin-like behavior, are smeared [see Appendix C].

**V. Superparaelectricity (SPE) and the conditions of its experimental observation**

Overlap of filled regions from Fig.3 ($T_f(R)<T<T_{cr}(R)$) and corresponding radii range ($R_{cr}<R<R_c$) fromFig.4 gives SPE phase region as shown in Fig.7 for PZT (a) and TGS (b).

At low temperatures $T \to 0$ (and so $R_{cr}(T) \to R_{cr}^0$), approximate expression for the freezing temperature $T_f$ could be obtained from the condition $\gamma(T_f, R \to R_{cr}^0) = 1$ (see Eq.(12)), namely we obtained the parabolic law:

$$T_f\left(R \to R_{cr}^0\right) \approx \frac{\pi R_{cr}^0 \cdot C_0}{3 k_B a_{11}} \left(\alpha_T T_c \left(R - R_{cr}^0\right)\right)^2 . \tag{20}$$

Constant $C_0$ depends on coefficients $a_{11}$ and $a_{12}$ only, as given by Eq.(12) for $\gamma$.

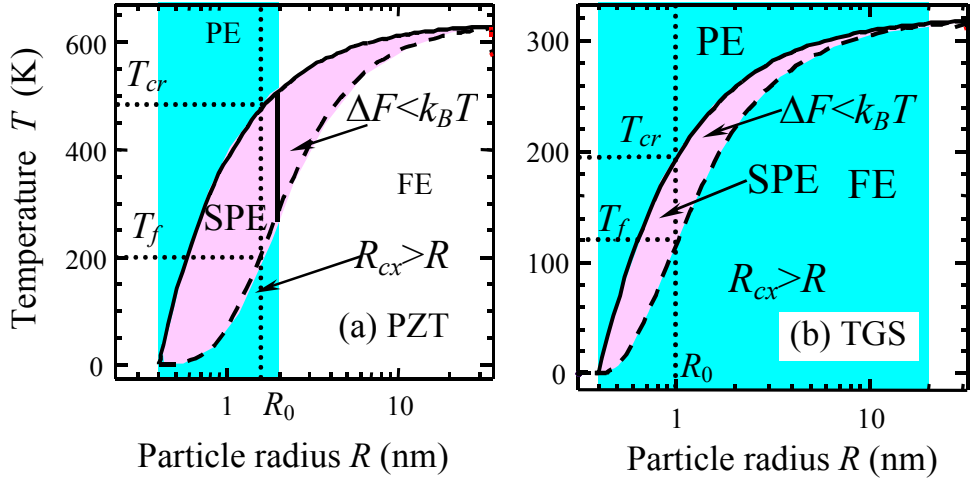

**Figure.7.** (Color online) SPE phase region for PZT (a) and TGS (b). Material parameters are listed in Fig.3.

To summarize, let's formulate the conditions of superparaelectric phase appearance in the ensemble of noninteracting ferroelectric particles of spherical shape and their properties peculiarities, which can be considered as characteristic features of SPE phase.

- The superparaelectric phase can appear in ferroelectric nanoparticles of average radius $R_{cr} < R < R_c$ at temperatures $T_f(R) < T < T_{cr}(R)$. In this region

(a) all nanoparticles dipole moments are aligned due to the correlation effects;

(b) potential barrier of polarization reorientation is smaller than the thermal activation energy $\sim k_B T$;

(c) Langevin-like law for polarization dependence on external field is valid at temperatures higher than the freezing temperature $T_f(R)$, but lower than the temperature $T_{cr}(R)$ of size-driven ferroelectric-paraelectric phase transition;

(d) hysteresis loop and remnant polarization (frozen SPE) appear at temperatures $T < T_f(R)$.

- The favorable conditions for the superparaelectricity observation in small ferroelectric nanoparticles at room temperatures are small Curie-Weiss constants and high nonlinear coefficients $a_{11}$ and narrow distribution function of particles radii. The ensemble of noninteracting ferroelectric nanoparticles could be realized in nanoporous nonferroelectric matrix with the porous filled at least partly by some ferroelectric material. Another type of composite material with cylindrical geometry of nanoporous in the nonferroelectric matrix filled with ferroelectric nanorods cannot be excluded also. However some difference in polar behavior for this geometry in comparison with spherical case can be expected.

The theoretical forecast is waiting for experimental revealing.

**Appendix A. Free energy expansion coefficients of order-disorder ferroelectrics**

Let us consider the order-disorder ferroelectrics.[26, 27] Neglecting the effects of tunneling and considering the dipoles with two possible orientations Hamiltonian of the system in Ising model can be written as follows:[28, 29]

$$H = -\frac{1}{2}\sum_{i,j} J_{ij} l_i l_j - \frac{1}{4}\sum_{i,j,k,m} J_{ijkm} l_i l_j l_k l_m - \sum_i p E_i l_i \quad (A.1)$$

Here axis z is pointed along dipole moments and indexes "z" are omitted for the simplicity. $J_{ij}$ and $J_{ijkm}$ are potentials of the two- and four-dipoles interactions respectively, $l_i$, $l_j$, $l_k$ and $l_m$ are z-components of the unit vectors pointed along electric dipoles ($l_{i,j,k,m}=\pm 1$), $p$ is dipole moment, $E_i$ is the z-component of the electric field acting on the dipole $l_i$. Summation in Eq. (A.1) on the indexes $i, j, k$ and $m$ is carried out on dipoles sites. With the help of Hamiltonian (A.1) in the molecular field approximation,[11] one can obtain the following free energy for the system of dipoles (see e.g. Refs. 28, 11):

$$F = -\frac{1}{2}\sum_j J_{ij}\langle l_i\rangle\langle l_j\rangle - \frac{1}{4}\sum_{i,j,k,m} J_{ijkm}\langle l_i\rangle\langle l_j\rangle\langle l_k\rangle\langle l_m\rangle - \sum_i p E_i \langle l_i\rangle + \\ + k_B T \sum_i \left(\frac{1+\langle l_i\rangle}{2}\ln\left(\frac{1+\langle l_i\rangle}{2}\right) + \frac{1-\langle l_i\rangle}{2}\ln\left(\frac{1-\langle l_i\rangle}{2}\right)\right) \quad (A.2)$$

Here $\langle l_i\rangle$ determines thermally averaged dipole moment of dipole "$i$" or the order parameter distribution across the system. The last sum represents entropy of the system. Minimization of Eq. (A.2) leads to the following equation for the thermally averaged moments $\langle l_i\rangle$:

$$-\frac{1}{2}\sum_j J_{ij}\langle l_j\rangle - \frac{1}{4}\sum_{j,k,m} J_{ijkm}\langle l_j\rangle\langle l_k\rangle\langle l_m\rangle - p E_i + k_B T \operatorname{arctanh}(\langle l_i\rangle) = 0 \quad (A.3)$$

Using the continuous approximation $\langle l_i\rangle \equiv l(\vec{r}_i)$ (see e.g. Refs. 28, 11) and the relationship $l(\vec{r}\pm\vec{a}) = l(\vec{r}) \pm (\vec{a}\nabla) l(\vec{r}) + (\vec{a}\nabla)^2 l(\vec{r})/2 \pm \ldots$ (here $\vec{a}$ is the vector determining the position of dipole) it is easy to write the equation determining the spatial distribution of order parameter:

$$-J l(\vec{r}) - J\delta\Delta l(\vec{r}) - J_{nl} l(\vec{r})^3 - p E + k_B T \operatorname{arctanh}(l(\vec{r})) = 0 \quad (A.4)$$

Here $J$ is effective interaction constant (the mean field acting on the each dipole from its neighbors), $J_{nl}$ is effective nonlinearity coefficient, $\delta$ determines the correlation between dipole moments. It can be easily shown, that the equation of state (A.4) corresponds to the minimum of the following free energy:

$$F = \iiint \begin{pmatrix} -\dfrac{J}{2}l^2(\vec{r}) - \dfrac{J_{nl}}{4}l^4(\vec{r}) - p\,E\,l(\vec{r}) - \dfrac{J\delta}{2}(\nabla l(\vec{r}))^2 + \\ + k_B T \left( \dfrac{1+l(\vec{r})}{2}\ln\left(\dfrac{1+l(\vec{r})}{2}\right) + \dfrac{1-l(\vec{r})}{2}\ln\left(\dfrac{1-l(\vec{r})}{2}\right) \right) \end{pmatrix} dV. \quad \text{(A.5a)}$$

Free energy (A.5a) can be easily expanded in series of $l(\vec{r})$ near zero up to sixth order:

$$F \approx \iiint \begin{pmatrix} \dfrac{k_B T - J}{2}l^2(\vec{r}) + \left(\dfrac{k_B T}{12} - \dfrac{J_{nl}}{4}\right)l^4(\vec{r}) + \dfrac{k_B T}{30}l^6(\vec{r}) + \ldots - \\ -p\,E\,l(\vec{r}) - \dfrac{J\delta}{2}(\nabla l(\vec{r}))^2 \end{pmatrix} dV. \quad \text{(A.5b)}$$

The temperature dependent terms of Eq. (A.5b) come from entropy part of the free energy (A.5a) (last term), all of them are positive. It is obvious that this approximation is not valid when values of $l(\vec{r})$ approaches unity. However, in the bulk system order parameter approaches unity only at high electric field and when temperature approaches absolute zero. Thus, not very far from phase transition point and for not very high electric fields one can use the more convenient algebraic form (A.5b).

**Appendix B. Depolarization field and correlation radius of a spherical particle**

*B.1. Depolarization field calculations in **r**-domain*

For depolarization field calculations we consider the case $E_0=0$. Electrostatic depolarization field is completely screened by the ambient free charges σ outside the particle: $\varphi_d(r=R)=0$ and $\varphi_d(r=0)$ is finite. The potential $\varphi_d(\mathbf{r})$ is nonzero inside the particle due to inhomogeneous polarization distribution depolarization [see Fig.1b, where the arrows lengths, indicating the absolute value of dipole moment, are different in different points inside the particle].

Maxwell's equation $\mathrm{div}(\mathbf{P}(\mathbf{r}) + \varepsilon_0 \mathbf{E}(\mathbf{r})) = 0$ for the inner electric field $\mathbf{E}_d = -\nabla \varphi_d(\mathbf{r})$, expressed via electrostatic potential $\varphi_d(\mathbf{r})$, and polarization $\mathbf{P}(\mathbf{r})$ has the form:

$$\Delta \varphi_d(\mathbf{r}) = \dfrac{\mathrm{div}\,\mathbf{P}(\mathbf{r})}{\varepsilon_0}. \quad \text{(B.1)}$$

$\varepsilon_0$ is the dielectric constant. In spherical coordinates $\mathbf{r} = \{r, \theta, \varphi\}$ the solution Eq.(B.1) acquires the form

$$\varphi_d(\mathbf{r}) = \begin{pmatrix} \dfrac{R}{4\pi}\int_0^{2\pi} d\varphi' \int_0^{\pi} d\theta' \dfrac{\sin\theta'(R^2 - r^2)f(\theta',\varphi')}{(r^2 + R^2 - 2Rr\cos\vartheta)^{3/2}} - \\ -\dfrac{1}{4\pi}\int_0^{2\pi} d\varphi' \int_0^{\pi} d\theta' \int_0^{R} dr' \dfrac{\sin\theta' \cdot r'^2\,\mathrm{div}\mathbf{P}(\mathbf{r}')}{\varepsilon_0 \sqrt{r^2 + r'^2 - 2r'r\cos\vartheta}} \end{pmatrix}. \quad \text{(B.2)}$$

Where $\cos[\vartheta(\theta,\theta',\varphi-\varphi')] = \cos\theta\cos\theta' + \sin\theta\sin\theta'\cos(\varphi-\varphi')$, function $f(\theta',\varphi')$ should be determined from the condition $\varphi_d(r=R,\theta,\varphi)=0$. Thus, we immediately obtained

$$f(\theta,\varphi) = \frac{1}{4\pi}\int_0^{2\pi}d\varphi'\int_0^{\pi}d\theta'\int_0^{R}dr'\frac{\sin\theta'\cdot r'^2\,\text{div}\mathbf{P}(\mathbf{r}')}{\varepsilon_0\sqrt{R^2+r'^2-2r'R\cos\vartheta}} \qquad (B.3)$$

At the surface: $\dfrac{1}{|\mathbf{r}-\mathbf{r}'|} = \dfrac{1}{\sqrt{r^2+r'^2-2r'r\cos\vartheta}} = \dfrac{1}{R}\sum_{n=0}^{\infty}\left(\dfrac{r'}{R}\right)^n P_n(\cos\vartheta)$.

Finally, Eqs.(B.2-3) can be rewritten as

$$\begin{cases} \varphi_d(r,\theta,\varphi) = \dfrac{R}{4\pi}\int_0^{2\pi}d\varphi'\int_0^{\pi}d\theta'\dfrac{\sin\theta'(R^2-r^2)g(\theta',\varphi',R)}{(r^2+R^2-2Rr\cos\vartheta)^{3/2}} - g(\theta,\varphi,r), \\ g(\theta,\varphi,r) = \dfrac{1}{4\pi}\int_0^{2\pi}d\varphi'\int_0^{\pi}d\theta'\int_0^{R}dr'\dfrac{\sin\theta'\cdot r'^2\,\text{div}\mathbf{P}(\mathbf{r}')}{\varepsilon_0\sqrt{r^2+r'^2-2r'r\cos\vartheta}}. \end{cases} \qquad (B.4)$$

*B.2. Depolarization field calculations in **k**-domain*

Partial inhomogeneous solution of Eq.(B.1) can be found for arbitrary fluctuations with the help of Fourier transformation. Namely, the Fourier integrals

$$\varphi_d(\mathbf{r}) = \frac{1}{(2\pi)^3}\int_{-\infty}^{\infty}\int_{-\infty}^{\infty}\int_{-\infty}^{\infty}dk_1\,dk_2\,dk_3\,\exp(i\mathbf{r}\mathbf{k})\widetilde{\varphi}(\mathbf{k}), \qquad (B.5a)$$

$$\mathbf{P}(\mathbf{r}) = \frac{1}{(2\pi)^3}\int_{-\infty}^{\infty}\int_{-\infty}^{\infty}\int_{-\infty}^{\infty}dk_1\,dk_2\,dk_3\,\exp(i\mathbf{r}\mathbf{k})\cdot\widetilde{\mathbf{P}}(\mathbf{k}), \qquad (B.5b)$$

$$\mathbf{E}(\mathbf{r}) = \frac{1}{(2\pi)^3}\int_{-\infty}^{\infty}\int_{-\infty}^{\infty}\int_{-\infty}^{\infty}dk_1\,dk_2\,dk_3\,\exp(i\mathbf{r}\mathbf{k})\cdot\widetilde{\mathbf{E}}(\mathbf{k}), \qquad (B.5c)$$

should be substituted into electrostatic Maxwell equations $\text{div}(\varepsilon_0\mathbf{E}+\mathbf{P})=0$, $\text{rot}(\mathbf{E})=0$. Then one obtains the following system of algebraic equations $\varepsilon_0(\mathbf{k}\cdot\widetilde{\mathbf{E}}(\mathbf{k}))+(\mathbf{k}\cdot\widetilde{\mathbf{P}}(\mathbf{k}))=0$, $[\mathbf{k}\times\widetilde{\mathbf{E}}(\mathbf{k})]=0$. The latter can be solved as $\widetilde{\mathbf{E}}(\mathbf{k})=\mathbf{k}\,A(\mathbf{k}) \Rightarrow \varepsilon_0(\mathbf{k}\cdot\mathbf{k})A(\mathbf{k})+(\mathbf{k}\cdot\widetilde{\mathbf{P}}(\mathbf{k}))=0 \Rightarrow A(\mathbf{k})=-\dfrac{(\mathbf{k}\cdot\widetilde{\mathbf{P}}(\mathbf{k}))}{\varepsilon_0 k^2}$, which gives the relationship between Fourier images of polarization and depolarization field:

$$\widetilde{\varphi}_p(\mathbf{k}) = -\frac{(\mathbf{k}\cdot\widetilde{\mathbf{P}}(\mathbf{k}))}{\varepsilon_0 k^2}, \qquad \widetilde{\mathbf{E}}(\mathbf{k}) = -\frac{\mathbf{k}(\mathbf{k}\cdot\widetilde{\mathbf{P}}(\mathbf{k}))}{\varepsilon_0 k^2}. \qquad (B.6)$$

Thus, the general inhomogeneous solution of Eq.(B.7) can be obtained as

$$\varphi_d(\mathbf{r}) = \int_{-\infty}^{\infty}\int_{-\infty}^{\infty} dk_1\, dk_2\, \exp(i x k_1 + i y k_2) \left( \begin{array}{l} \exp\!\left(-z\sqrt{k_1^2+k_2^2}\right) A(k_1,k_2) + \exp\!\left(z\sqrt{k_1^2+k_2^2}\right) B(k_1,k_2) \\ - \int_{-\infty}^{\infty} dk_3\, \exp(i z k_3) \dfrac{(\mathbf{k}\cdot \widetilde{\mathbf{P}}(\mathbf{k}))}{\varepsilon_0 k^2} \end{array} \right) \quad (B.7)$$

Functions $A$ and $B$ should be found from the conditions $\varphi_d(R)=0$ and finiteness of $\varphi_d(0)$. Direct variation of Eq.(B.7) under the condition of fixed variations $\delta P_3(R) + \lambda\, d\delta P_3(R)/dr = 0$ leads to the relation between depolarization field z-component $E_d$ and polarization $\delta P_3(\mathbf{r})$ in Fourier representation:

$$\widetilde{E}_d[\delta P_3] = -\frac{k_3^2}{\varepsilon_0 k^2}\,\delta\widetilde{P}_3(\mathbf{k}). \quad (B.8)$$

*B.3. Correlation radius of nanoparticle*

Equation for the Green function of Eq.(5a) has the form

$$\left(a_1 + 3 a_{11} P_0^2(\mathbf{r})\right) G(\mathbf{r},\mathbf{r}') + E_d[G(\mathbf{r},\mathbf{r}')] - \zeta \frac{\partial^2 G(\mathbf{r},\mathbf{r}')}{\partial z^2} - \eta\!\left(\frac{\partial^2 G(\mathbf{r},\mathbf{r}')}{\partial x^2} + \frac{\partial^2 G(\mathbf{r},\mathbf{r}')}{\partial y^2}\right) = \delta(\mathbf{r}-\mathbf{r}') \quad (B.9)$$

Here $\delta(\mathbf{r}-\mathbf{r}')$ is Dirac delta-function. Using the integral representation of Dirac delta-function[30]

$$\delta(\mathbf{r}-\mathbf{r}') = \frac{1}{(2\pi)^2} \int_{-\infty}^{\infty}\int_{-\infty}^{\infty}\int_{-\infty}^{\infty} dk_1\, dk_2\, dk_3\, \exp(i(\mathbf{r}-\mathbf{r}')\mathbf{k}),$$

the solution of Eq. (B.9) allowing for Eqs.(5b) and (9) in the **k**-domain has the form

$$\widetilde{G}(\mathbf{k}) \approx \frac{1}{3 a_{11}\overline{P}_0^2 + a_R(T,R) + \eta(k_1^2 + k_2^2) + \left(\zeta + 1/\varepsilon_0 k^2\right) k_3^2}. \quad (B.10)$$

Approximation in (B.10) is related with substitution $P_0^2(\mathbf{r}) \to \overline{P}_0^2$, depolarization field and surface influence. The renormalization of coefficient $a_1$:

$$a_1 \;\to\; a_R(T,R) \approx \alpha_T\!\left(T - T_c\!\left(1 - \frac{R_{cr}^0}{R}\right)\right). \quad (B.11)$$

Where $R_{cr}^0 = \dfrac{3\xi}{\alpha_T T_c\left(\lambda + \sqrt{\xi\varepsilon_0}\right)}$ is a critical radius at zero temperature. At a given temperature $T$ the sphere critical radius $R_{cr}(T) \approx \dfrac{R_{cr}^0}{1 - T/T_c}$, existing at $T<T_c$, should be found from the condition $a_R(T,R_{cr})=0$. The equilibrium polarization $\overline{P}_0$ averaged over the nanoparticle volume satisfies the equation $a_R(T,R)\overline{P}_0 + a_{11}\overline{P}_0^3 = E_0$. At zero external field $E_0 = 0$, the spontaneous polarization is nonzero in ferroelectric phase, $\overline{P}_0^2 = -a_R(T,R)/a_{11}$, while $\overline{P}_0^2 = 0$ in paraelectric phase.

Using the relation $G(\mathbf{r},\mathbf{r}') = k_B T \chi(\mathbf{r},\mathbf{r}')$, the expressions (B.10) can be simplified to simple Lorenzian form

$$\tilde{\chi}(k_1=0, k_2=0, k_3) = \begin{cases} \dfrac{1}{(\varepsilon_0^{-1} + a_R(T,R))(1+R_{cz}^2 k_3^2)}, & a_R(T,R) > 0, \\ \dfrac{1}{(\varepsilon_0^{-1} - 2a_R(T,R))(1+R_{cz}^2 k_3^2)}, & a_R(T,R) < 0. \end{cases} \quad \text{(B.12a)}$$

$$\tilde{\chi}(k_1, k_2, k_3=0) = \begin{cases} \dfrac{1}{a_R(T,R)(1+R_{cx}^2(k_1^2+k_2^2))}, & a_R(T,R) > 0, \\ \dfrac{-1}{2a_R(T,R)(1+R_{cx}^2(k_1^2+k_2^2))}, & a_R(T,R) < 0. \end{cases} \quad \text{(B.12b)}$$

Where the correlation radii for the fluctuations along and across polar axis are introduced as:

$$R_{cz}(T,R) = \begin{cases} \sqrt{\dfrac{\xi}{\varepsilon_0^{-1} + a_R(T,R)}}, & a_R(T,R) > 0 \\ \sqrt{\dfrac{\xi}{\varepsilon_0^{-1} - 2a_R(T,R)}}, & a_R(T,R) < 0 \end{cases}, \quad \text{(B.13a)}$$

$$R_{cx}(T,R) = \begin{cases} \sqrt{\dfrac{\eta}{a_R(T,R)}}, & a_R(T,R) > 0 \\ \sqrt{\dfrac{-\eta}{2a_R(T,R)}}, & a_R(T,R) < 0 \end{cases}. \quad \text{(B.13b)}$$

Bulk correlation radiuses are $R_{cx}^b = \sqrt{-\eta/2a_1(T)}$ and $R_{cz}^b = \sqrt{\xi/(\varepsilon_0^{-1} - 2a_1(T))}$ in ferroelectric phase ($T < T_c$).

**Appendix C. Polarization response calculations**

*C.1. Potential barrier for perovskite ferroelectric nanoparticles*

Thermodynamic free energy expansion for spherical ferroelectric particle with cubic symmetry has the form:

$$F[\overline{\mathbf{P}}] = V \left( \begin{array}{l} \dfrac{a_R(T,R)}{2}(P_1^2 + P_2^2 + P_3^2) + \dfrac{a_{12}}{2}(P_1^2 P_2^2 + P_3^2 P_2^2 + P_3^2 P_1^2) + \\ + \dfrac{a_{11}(T,R)}{4}(P_1^4 + P_2^4 + P_3^4) - P_1 E_{01} - P_2 E_{02} - P_3 E_{03} \end{array} \right) \quad \text{(C.1)}$$

In Eq.(C.1) the integration over the particle volume $V = 4\pi R^3/3$ was performed. Free energy (C.1) is stable at high polarization values only for $a_{11} > 0$ and $a_{11} + 2a_{12} > 0$, otherwise one should consider higher order terms in Eq. (C.1). At zero external electric field, $E_0 = 0$, free energy (C.1) can describe three different phases.

(a) <u>PE-phase</u>. At $a_R > 0$ only paraelectric phase (PE) exists with zero polarization $P_i = 0$. Corresponding free energy density is zero.

(b) <u>rFE-phase</u>. For the case $a_R < 0$ and $a_{11} > a_{12}$ only rhombohedral ferroelectric phase (rFE) is thermodynamically stable with polarization components equal to $P_1^2 = P_2^2 = P_3^2 = \dfrac{-a_R}{a_{11} + 2a_{12}}$.

Corresponding free energy $g = F/V$ density is $\dfrac{-3a_R^2}{4(a_{11} + 2a_{12})}$.

(c) <u>tFE-phase</u>. For the case $a_R < 0$ and $a_{11} < a_{12}$ only tetragonal ferroelectric phase (tFE) is thermodynamically stable with polarization components equal to $P_i^2 = -a_R/a_{11}$, $P_j^2 = P_k^2 = 0$

(i,j,k=1,2,3). Corresponding free energy density is $g = \dfrac{-a_R^2}{4a_{11}}$. At $a_R < 0$ the saddle points

$P_i^2 = P_j^2 = \dfrac{-a_R}{a_{11} + a_{12}}$, $P_k^2 = 0$ exist. They border the minima of stable phase. Corresponding

density of free energy (C.1) is $g = \dfrac{-a_R^2}{2(a_{11} + a_{12})}$.

One can find the potential barrier between different polarization orientations in stable phases as a difference between the free energy values corresponding to minimum and saddle point. For the rFE-phases and tFE-phases reorientation potential barrier has the form

$$\Delta F = V \begin{cases} \dfrac{a_R^2(T,R)}{4a_{11}} \left( \dfrac{a_{12} - 2a_{11}}{a_{12} + 2a_{11}} \right), & 2a_{11} < a_{12} \\ \dfrac{a_R^2(T,R)}{4a_{11}(1 + a_{12}/a_{11})} \left( \dfrac{2a_{11} - a_{12}}{a_{12} + 2a_{11}} \right), & 2a_{11} > a_{12}. \end{cases} \quad (C.2)$$

*C.2. Polarization response calculations*

At a given temperature $T$ and nanoparticle radius $R$, polarization orientation with respect to external field $\mathbf{E}_0 = \{0, 0, E_0\}$ can be written as $\mathbf{P} = \{P \sin\theta \cos\varphi, P \sin\theta \sin\varphi, P \cos\theta\}$. Thus, the free energy (11) acquires the form:

$$F(\overline{P}, \theta, \varphi) = V \begin{cases} \left( \dfrac{a_R(T,R)}{2} \overline{P}^2 + \dfrac{a_{12}}{2} \overline{P}^4 (\sin^4\theta \sin^2\varphi \cos^2\varphi + \sin^2\theta \cos^2\theta) \\ + \dfrac{a_{11}}{4} \overline{P}^4 (\sin^4\theta(\cos^4\varphi + \sin^4\varphi) + \cos^4\theta) - E_0 \overline{P} \cos\theta \right), & \text{perovskite FE,} \\ \left( \dfrac{a_R(T,R)}{2} \overline{P}^2 + \dfrac{a_{11}}{4} \overline{P}^4 - E_0 \overline{P} \cos\theta \right), & \text{uniaxial FE.} \end{cases}$$

(C.3)

Averaging on the angle φ leads to:

$$\int_0^{2\pi} d\varphi \exp\left(-\frac{F(P,\theta,\varphi)}{k_B T}\right) = 2\pi \exp\left(-\frac{F(P,\theta,0)}{k_B T} - \frac{a_{11}P^4}{16k_B T}\left(\frac{a_{12}}{a_{11}} - 1\right)\sin^4\theta\right) I_0\left(\frac{a_{11}P^4}{16k_B T}\left(\frac{a_{12}}{a_{11}} - 1\right)\sin^4\theta\right)$$

$$\approx 2\pi\left(1 + \pi\frac{a_{11}P^4}{8k_B T}\left(\frac{a_{12}}{a_{11}} - 1\right)\sin^4\theta\right)^{-1/2} \exp\left(-\frac{F(P,\theta,0)}{k_B T}\right)$$

(C.4)

Relation between distribution functions: $\tilde{\mu}(R) = \mu(R)\frac{dP}{dR}$.

$$T_f\left(R \to R_{cr}^0\right) \approx \frac{V}{4k_B a_{11}}\left(\alpha_T T_c\left(1 - \frac{R_{cr}^0}{R}\right)\right)^2 \begin{cases} \dfrac{a_{12} - 2a_{11}}{a_{12} + 2a_{11}}, & 2a_{11} < a_{12}, \quad \text{perovskite FE}, \\ \dfrac{a_{11}}{a_{11} + a_{12}}\left(\dfrac{2a_{11} - a_{12}}{a_{12} + 2a_{11}}\right), & 2a_{11} > a_{12}, \quad \text{perovskite FE}, \\ 1, & \text{uniaxial FE}. \end{cases}$$